\documentclass[prd,nofootinbib,preprint,superscriptaddress]{revtex4}

\usepackage{amsmath, amssymb, amsthm, graphicx, epsfig, fancyhdr,epsfig, slashed}

\DeclareUnicodeCharacter{202F}{\,}

\usepackage{tikzsymbols}
\usepackage{natbib}
\usepackage{float}
\usepackage{xcolor}

\usepackage[utf8]{inputenc}
\usepackage[T1]{fontenc}

\usepackage{enumitem}

\usepackage{tikz,xcolor,hyperref}

\usepackage{slashed}
\usepackage{braket}

\usepackage{hhline}
\usepackage{multirow}

\definecolor{lime}{HTML}{A6CE39}
\DeclareRobustCommand{\orcidicon}{
	\begin{tikzpicture}
	\draw[lime, fill=lime] (0,0) 
	circle [radius=0.2] 
	node[white] {{\fontfamily{qag}\selectfont \tiny ID}};
	\draw[white, fill=white] (-0.0625,0.095) 
	circle [radius=0.007];
	\end{tikzpicture}
	\hspace{-2mm}
}

\foreach \x in {A, ..., Z}{\expandafter\xdef\csname orcid\x\endcsname{\noexpand\href{https://orcid.org/\csname orcidauthor\x\endcsname}
			{\noexpand\orcidicon}}
}

\newcommand{\be}{\begin{equation}}
\newcommand{\ee}{\end{equation}}
\newcommand{\bea}{\begin{eqnarray}}
\newcommand{\eea}{\end{eqnarray}}

\newcommand{\ba}{\begin{eqnarray}}
\newcommand{\ea}{\end{eqnarray}}
\newcommand{\bi}{\begin{itemize}}
\newcommand{\ei}{\end{itemize}}

\newcommand{\x}{\star}

\DeclareUnicodeCharacter{2212}{-}

\begin{document}

\title{Testing Seesaw and Leptogenesis Gravitational Waves:\\ \it{Majorana versus Dirac cases}\footnote{This paper is the original version of the article accepted for publication Phys.Rev.D with the title “Imprints of seesaw and leptogenesis on cosmic-string contributions to the stochastic gravitational wave background:  \\ \it{Majorana versus Dirac cases}”}}


\author{Anish Ghoshal}
\email{anish.ghoshal@fuw.edu.pl}
\affiliation{Department of Physics and Astronomy, University of Sussex, \\
Brighton, BN1 9RH, United Kingdom}

\author{Kazunori Kohri}
\email{kazunori.kohri@gmail.com}
\affiliation{Division of Science, National Astronomical Observatory of Japan, Mitaka, Tokyo 181-8588, Japan}
\affiliation{The Graduate University for Advanced Studies (SOKENDAI), Mitaka, Tokyo 181-8588,
Japan}
\affiliation{Theory Center, IPNS, KEK, 1-1 Oho, Tsukuba, Ibaraki 305-0801, Japan}
\affiliation{Kavli IPMU (WPI), UTIAS, The University of Tokyo, Kashiwa, Chiba 277-8583, Japan.}

\author{Nimmala Narendra}
\email{nimmalanarendra@gmail.com}
\affiliation{Department of Physics, PES Institute of Technology $\&$ Management, Sagar Road, Shivamogga, Karnataka-577204, India}

\begin{abstract}
We investigate the B-L gauge extension of the Standard Model that the Dirac seesaw mechanism with thermal Leptogenesis leave imprints in early universe via the stochastic gravitational background (SGWB) emanating from a network of cosmic strings when B-L symmetry is broken. With right-handed neutrino mass lighter than the typical scale of grand unification, the B-L symmetry protecting the right-handed neutrinos leads to constraints on the Yukawa couplings for both Dirac and Majorana scenarios. Estimating the predicted gravitational wave background we find that future space-borne missions could probe the range concerning thermal Dirac Leptogenesis. In a comparative analysis between such probes of gravitational wave sourced from cosmic strings in Dirac and Majorana Leptogenesis in the B-L extension, based on the energy scales of the Leptogenesis, for instance, GW detectors will be able to probe the scale of Dirac Leptogenesis upto $ 10^{9}$ GeV, while for Majorana Leptogenesis it would be upto $ 10^{12}$ GeV. 
\end{abstract}

\maketitle
\tableofcontents
\flushbottom

\section{Introduction}\label{sec:intro}



The evidence of going beyond the Standard Model (SM) of particle physics arises undisputedly from neutrino oscillation experiments involving solar \cite{Super-Kamiokande:2001bfk,Super-Kamiokande:2002ujc,SNO:2002tuh,Super-Kamiokande:2005mbp,Super-Kamiokande:2016yck,Borexino:2015axw}, atmospheric \cite{IceCube:2017lak, ANTARES:2018rtf} and reactor \cite{KamLAND:2008dgz,T2K:2011ypd,DoubleChooz:2011ymz,T2K:2013ppw}. We infer from these experiments that neutrinos 
have mass and the flavor states mix due to the propagation of multiple mass eigenstates. This however only quantifies the squared mass differences of the neutrinos and not the absolute mass scales involved. On the other hand, the $\beta$-decay experiment KATRIN \cite{KATRIN:2021uub} gives a stringent direct limit on the absolute value of neutrino mass scale, $m_\nu < 0.8 $ eV.

On the cosmological frontiers, measurements of the Cosmic Microwave Background radiation (CMBR) by Planck 2018 \cite{Planck:2018vyg} and large-scale structure (LSS) constrain the sum of all neutrino masses to $\sum_i m_{\nu_i} < 0.12$ eV \cite{Planck:2018vyg,eBOSS:2020yzd}. On the other hand, the observed baryon asymmetry of the Universe (BAU) \cite{Zyla:2020zbs,Planck:2018vyg}, often expressed in terms of the  baryon to photon ratio is, according to the Planck data~\cite{Planck:2018vyg} \footnote{It is also expressed as $Y_B = \frac{n_{\bar{B}}}{s} =
 (0.82 - 0.92) \times 10^{-10}$. We use the central value $Y_B = \frac{n_{\bar{B}}}{s} =
 8.66 \times 10^{-11}$ in our estimates.}, 
\begin{equation}
\eta_B^{\rm CMB} = \frac{n_{B}-n_{\bar{B}}}{n_{\gamma}} = (6.21 \pm 0.16) \times 10^{-10}
\label{eq:eta}
\end{equation}
This agrees with the values of abundances of light elements extracted from BBN  (Big Bang Nucleosynthesis) data~\cite{Fields:2019pfx}. 

Just augmenting the Standard Model with two or more right-handed Majorana neutrinos (RHNs), one can explain the tiny neutrino masses, generated via the well-known Type-I seesaw mechanism \cite{Minkowski:1977sc,Yanagida:1979as,Glashow:1979nm,Mohapatra:1979ia}, 
while baryon asymmetry can be explained using the baryogenesis via Leptogenesis mechanism~\cite{Fukugita:1986hr}, provides a unified explanation. Nonetheless this mechanism demands the lightest RHN mass to exceed the Davidson-Ibarra bound\cite{Davidson:2002qv}, $M_1 \gtrsim 10^9$ GeV, assuming the RHN mass spectrum is hierarchical, along with no lepton flavor effects~\cite{Nardi:2006fx,Abada:2006fw,Abada:2006ea,Giudice:2003jh}. This deflates any hope of direct detection of the high-scale seesaw and Leptogenesis since they lie well beyond the energy reaches of the current and future laboratory experiments. 

Nonetheless, certain indirect signatures of new physics like lepton number violation processes through neutrinoless double beta decay~\cite{Cirigliano:2022oqy} or CP violation in neutrino oscillation~\cite{Endoh:2002wm} are searched for in laboratories. There can be certain theoretical constraints on the low energy values of couplings, whose consistency conditions for UV-completion such as in $SO(10)$ Grand Unified Theories (GUT) ~\cite{DiBari:2008mp,Bertuzzo:2010et,Buccella:2012kc,Altarelli:2013aqa,Fong:2014gea,Mummidi:2021anm,Patel:2022xxu} or with the demand of electroweak (EW) vacuum (meta)stability in the early Universe~\cite{Ipek:2018sai,Croon:2019dfw} sometimes provide some useful bounds on the large parameter space involved in Seesaw and Leptogenesis scenarios. 

The discovery of Gravitational Waves (GWs) from black hole mergers by LIGO, Virgo, KAGRA collaboration \cite{LIGOScientific:2016aoc,LIGOScientific:2016sjg} and evidence for measurements of stochastic GW background by pulsar timing array (PTA) observations~\cite{Carilli:2004nx,Janssen:2014dka,Weltman:2018zrl,EPTA:2015qep,EPTA:2015gke,NANOGrav:2023gor,NANOGrav:2023hvm} have lead to several new physics cases to detect GWs of primordial and cosmological origins as well. In the context of baryogenesis via Leptogenesis scenarios investigated so far, cosmological pathways to probe such high-scale physics involve predictions of the CMB spectral indices~\cite{Ghoshal:2022fud} or gravitational waves from local cosmic strings~\cite{Dror:2019syi, Saad:2022mzu, DiBari:2023mwu, Blasi:2020wpy}, global cosmic strings \cite{Fu:2023nrn}, domain walls~\cite{Barman:2022yos, King:2023cgv}, nucleating and colliding vacuum bubbles~\cite{Dasgupta:2022isg,Borah:2022cdx,Ghosh:2025non}, graviton bremmstrahlung~\cite{Ghoshal:2022kqp,Datta:2024tne,Datta:2025wfh}, inflationary tensor perturbations of first-order~\cite{Berbig:2023yyy,Borboruah:2024eal} and second-order~\cite{Afzal:2024xci,Chianese:2025mll}, primordial blackholes~\cite{Perez-Gonzalez:2020vnz,Datta:2020bht,JyotiDas:2021shi,Barman:2021ost,Bernal:2022pue,Bhaumik:2022pil,Bhaumik:2022zdd,Ghoshal:2023fno}, and utisilizing features of the primordial non-gaussianity and curvature bi-spectrum and tri-spectrum in CMB \cite{Cui:2021iie,Fong:2023egk}. Particularly, it has been shown that one may probe directly as a possible foorptint of high-scale leptogenesis with GW from cosmic strings, via investigating only the RHN mass sector, the origin of which may occur at a more fundamental level due to symmetry breaking~\cite{Dror:2019syi,Blasi:2020wpy,Samanta:2020cdk,Datta:2020bht,Chianese:2024gee}. For instance, the RHNs become massive because of the spontaneous breaking of a $U(1)_{B-L}$ gauge invariance, a symmetry which quite naturally gets embedded in several Grand Unified Theories (GUT) theories \cite{Davidson:1978pm,Marshak:1979fm,Buchmuller:2013lra,Buchmuller:2019gfy}, and a breaking of which inevitably produces cosmic string network that radiates gravitational wave signals. For any $U(1)_{B-L}$ invariant RHN mass term, the Yukawa coupling $y_N \overline{N_R} \Phi N^C_R$, where $y_N$ is the Yukawa coupling, $N_R$ represents the RHN particle and $\Phi$ denotes the $B-L$ scalar with vacuum expectation value $v_\Phi$. When the $U(1)_{B-L}$ symmetry is broken, the RHNs obtain mass given by $M_N=y_N v_\Phi$, thereby relating to the properties (like string tension) of cosmic strings which are formed after the symmetry breaking~\cite{Kibble:1976sj,Hindmarsh:1994re,Jeannerot:2003qv} leading to production of GWs with a almost scale-invariant spectrum with an overall GW amplitude $\Omega_{\rm GW}\propto v_\Phi$~\cite{Vilenkin:1981bx,Vachaspati:1984gt}. 

Under this circumstance, while most of the Leptogenesis scenarios consider Majorana nature of light neutrinos, one equally appealing alternative is to consider Dirac nature of light neutrinos. As originally proposed in Refs.~\cite{Dick:1999je, Murayama:2002je}, one can have successful Leptogenesis even with light Dirac neutrino scenarios where total lepton number or $B-L$ is conserved just like it is conversed in the SM of particle physics \footnote{The recently proposed mechanism of wash-in Leptogenesis \cite{Domcke:2020quw,Domcke:2022kfs} advocates an approach getting rid of CP violation
in the RHN sector altogether and separates the energy
scales of CP violation and B-L violation altogether in presence of non-trivial chemical background configurations. Like Dirac Leptogenesis, here the RHNs mere spectator processes that reprocess the chemical potentials of the SM
particle species in the thermal bath. The action of the
electroweak sphalerons on the chemical composition of
the SM plasma then results in the usual violation of
baryon-plus-lepton number B+L, while the action of the
RHN interactions results in the violation of B-L, also see Refs.\cite{Maleknejad:2014wsa,Maleknejad:2020yys,Maleknejad:2020pec} for gravitational production during inflation. }. This mechanism is popularly known as the Dirac Leptogenesis. What this mechanism exploits is the creation of an equal and opposite amount of lepton asymmetry in left handed and right handed neutrino sectors  which is subsequently followed by the conversion of left sector asymmetry into baryon asymmetry via the electroweak sphalerons. This is made possible since the right-handed sector is invisible to the sphalerons inside the SM plasma, only the left-handed asymmetry is converted into baryons. The asymmetry within the $\nu_R$ sector, and indeed the $\nu_R$ particles themselves, are well-high impossible to observe due to their extreme weak interactions. Needless to mention here, the effective Dirac Yukawa couplings are naturally very tiny which guarantee that the lepton asymmetries in the left and right handed sectors do not equibriliate and washed out. A myriad of possible implementation of this idea have been considered, see some in Refs.~\cite{Boz:2004ga, Thomas:2005rs, Thomas:2006gr, Cerdeno:2006ha, Gu:2006dc, Gu:2007mc, Chun:2008pg, Bechinger:2009qk, Chen:2011sb, Choi:2012ba, Borah:2016zbd, Gu:2016hxh, Narendra:2017uxl}. Particularly, in a few related works~\cite{Heeck:2013vha, Gu:2019yvw, Mahanta:2021plx}, violation of $B-L$ symmetry was accommodated in a way to preserve the Dirac nature of light neutrinos, while generating lepton asymmetry simultaneously.

In this paper we first discuss Dirac Leptogenesis, with thermalized RHNs and therefore suitable within the neutrinophilic \cite{Ma:2000cc} two-Higgs-doublet which naturally explains the smallness of Dirac masses \cite{Wang:2006jy,Gabriel:2006ns,Davidson:2009ha,Davidson:2010sf}. Moreover we follow the framework of lepton-number-violating (LNV) Dirac neutrinos~\cite{Heeck:2013rpa,Heeck:2013vha} to create a lepton asymmetry from the $CP$-violating decay of a heavy particle that is present in the seesaw model.\footnote{Prior to Ref.~\cite{Heeck:2013rpa},  Ref.~\cite{Chen:2012jg} mentioned roles of LNV Dirac neutrinos in the early Universe.} Due to this reason, the mechanism is actually more close to standard Leptogenesis than neutrinogenesis, despite the fact that it contains Dirac neutrinos. Finally we do a detailed investigation of the parameter space of leptogenesis, in B-L extensions of the SM, comparing Majorana and Dirac leptogenesis. In terms of scales of seesaw and Leptogensis, we provide quantitative estimates regarding various GW detectors like LISA and ET which are able to measure GW from B-L symmetry breaking cosmic strings, probing large regions of such parameter space in consists of different correlations between seesaw scale and the scales of Dirac and Majorana leptogenesis. In the current scenario, we do not consider the lepton flavor effects \cite{Nardi:2006fx,Abada:2006fw,Abada:2006ea}, throughout the draft, for simplicity.

\textit{The paper is organized as follows:} In Sec.\,\ref{sec:model}\,, we give brief introduction to Dirac Leptogenesis model and in the following Sec.\,\ref{sec:dirac_lept}\,, we discuss the Dirac neutrinos and Dirac Leptogenesis. The origin of the Gravitational waves from the cosmic strings discussed in \,\ref{sec:GW_local_cs}\,. The numerical results while comparing the Dirac and Majorana Leptogenesis with gravitational wave signal are discussed in Sec.\,\ref{sec:Num_res}\,. Finally we conclude in Sec.\,\ref{sec:conclusion}\,.

\medskip

\section{The B-L extended Model} \label{sec:model}

We extend the SM symmetry with a well motivated beyond the SM framework based on the gauged $U(1)_{B-L}$ symmetry, where $B$ and $L$ are the baryon and lepton numbers respectively. Considering the gauged $U(1)_{B-L}$ symmetry brings unwanted triangular anomalies to the framework. These anomalies break gauge symmetry at the quantum level, caused by chiral fermions running in triangle loop diagrams involving gauge bosons.\footnote{ The triangle anomalies for both $U(1)^3_{B-L}$ and the mixed $U(1)_{B-L}-(\text{gravity})^2$ diagrams are non-zero. These triangle anomalies for the SM fermion content turns out to be $\mathcal{A}^{\text{SM}} \left[ U(1)^3_{B-L} \right]=-3$, $\mathcal{A}^{\text{SM}} \left[ (\text{gravity})^2 \times U(1)_{B-L} \right]=-3$.  } These anomalies can be canceled by introducing three right-handed neutrinos to the framework \footnote{ The right-handed neutrinos contribute $\mathcal{A}^{\text{New}} \left[ U(1)^3_{B-L} 
\right] = 3, \mathcal{A}^{\text{New}} \left[ (\text{gravity})^2 \times U(1)_{B-L} \right] = 3$ leading to vanishing total of triangle anomalies \cite{Borah:2018smz}.}, which is also essential for realizing the neutrino masses\cite{Geng:1989tcu, He:1990me, Kohri:2013sva, Narendra:2018vfw, Borah:2018smz}. The overall particle content of our model extend with three right-handed neutrinos $\nu_{R_\alpha},\, \alpha=\{1, 2, 3\}$, two singlet scalars $\varphi$ and $\xi$, under the SM gauge group. The scalar $\varphi$ is introduced in order to spontaneously break the $B-L$ symmetry by four units. The $\xi$ acts as a mediator which is responsible for the creation of asymmetry in the right-handed neutrinos \cite{Heeck:2013vha}.  The $B-L$ charge assignment of the new particle content can be seen in Table \ref{table}\,.

\begin{table}[h]
\begin{center}
\begin{tabular}{|c|c|c|c|}
\hline
Fields & $SU(2)_L$ & $U(1)_Y$ & $U(1)_\text{B-L}$  \\
\hline
$\nu_{R}$ & 1 & 0 &  -1      \\
$\varphi$  & 1 & 0 & 4  \\
$\xi$  & 1 & 0 & -2  \\
\hline
\end{tabular}
\end{center}
\caption{\it Charge assignment of the additional fields under the extended gauge symmetry. }
\label{table}
\end{table}

The relevant Lagrangian takes the following form:
\begin{eqnarray}
 \mathcal{L}  &=& \mathcal{L}_\mathrm{SM} +\mathcal{L}_\mathrm{kinetic} +\mathcal{L}_{Z'}- V(H,\varphi,\xi) \nonumber\\
&&\quad + \left( y_{\alpha\beta} \overline{L}_\alpha H \nu_{R,\beta} + \frac{1}{2}\kappa_{\alpha\beta}\, \xi \, \overline{\nu}_{R,\alpha} \nu_{R,\beta}^c + H.c.\right) ,
\label{eq:model}
\end{eqnarray}
where 
\begin{eqnarray}
 V(H,\varphi,\xi) &\equiv & \sum_{X= H, \varphi, \xi} \left(\mu_X^2 |X|^2 + \lambda_X |X|^4\right) \nonumber\\
 && \quad + \sum_{\substack{X, Y= H, \varphi, \xi\\ X\neq Y}}  \frac{\lambda_{X Y}}{2} |X|^2 |Y|^2
 - \mu\left( \varphi \,\xi^2 + H.c \right) ,
\label{eq:potential}
\end{eqnarray}
The $\mathcal{L}_\mathrm{SM},\,\mathcal{L}_\mathrm{kinetic},\,\mathcal{L}_{Z'}$ are the all SM particle interactions, kinetic terms of new particle content and $Z'$ boson interaction terms, respectively. $H$ is the SM Higgs doublet. We consider $\mu_{H}^{2},\,\mu_{\varphi}^{2}<0<\mu_{\xi}^2$. Hence the potential has minimum with $\langle\xi\rangle=0$, $\langle H \rangle\neq0\neq\langle\varphi\rangle$, which breaks the $SU(2)_L\times U(1)_Y\times U(1)_{B-L}$ to $U(1)_{\text{EM}}\times Z_{4}^L$. Under $Z_{4}^L$ the leptons transform as $\ell \rightarrow -i\,\ell$ and $\xi\rightarrow-\xi$. In principle the $\mu$ coupling in the last term in Eq. \ref{eq:potential} can induce a non-zero vev for $\xi$ after $\varphi$ acquires a vacuum expectation value (VEV). If $\langle \xi \rangle \neq 0$, it spontaneously breaks the residual symmetry $Z_4^L$, that would generate Majorana mass terms for $\nu_R$, which leads to a standard Majorana Leptogenesis scenario, rather than the intended Dirac Leptogenesis framework. Since $\xi$ does not acquire a VEV and forbids Majorana mass term it allows small Dirac neutrino mass $m_{\nu_{\alpha \beta}}= y_{\alpha \beta} \langle H \rangle$ with the Yukawa coupling of the order $ \lesssim 10^{-11}$. The Yukawa coupling $\kappa_{\alpha \beta}$ is a complex symmetric matrix, in general. The $\kappa_{\alpha \beta}$ plays a crucial role in generating the $\nu_R$ asymmetry.  We define $\xi=(\zeta_1+ i\, \zeta_2)/\sqrt{2}$, where $\zeta_{1,2}$ real scalars. The $\mu$ term present in the potential induces a mass-splitting between the real scalars $\zeta_{1,2}$:
\begin{equation}
M_1^2=M_c^2-2\mu \langle \varphi \rangle \,,\,\,~~~M_2^2=M_c^2+2\mu \langle \varphi \rangle \,,   \label{eq:m1m2_rel}
\end{equation}
where $\langle \varphi \rangle \equiv \eta$ is the vev of $\varphi$. The $M_c$ is a common mass term for the real scalars $\zeta_{1,2}$. The $M_c$ decides the mass scale of the $M_{\zeta_i}$. The $\mu$ is the coupling strength of $\varphi$ with $\xi$.

The mass scale of $Z_{B-L}$ boson is very heavy, therefore it decouples from the thermal bath early in the Universe, and hence it does not thermalizes the right-handed neutrinos. The gauge interactions mediated by heavy $Z_{B-L}$ boson or the scalar interactions involving $\varphi$ are not sufficient enough to thermalize the RHN.

\section{Dirac Neutrinos and Lepton Number Violating Dirac Leptogenesis}\label{sec:dirac_lept}
In our scenario the neutrinos are Dirac particles, yet the $B-L$ number is violated. A heavy scalar particle {\it i.e.,} $\xi$, generates an asymmetry in the right-handed neutrinos in its CP-violating out-of-equilibrium decay. To generate such an asymmetry we need a second copy of $\xi$, {\it i.e.,} $\xi_{1,2}$. The $\xi_{2}$ is defined as $\xi_2=(\zeta_3+ i\, \zeta_4)/\sqrt{2}$ and also have the mass splitting similar to Eq.\,\ref{eq:m1m2_rel},
\begin{equation}
M_3^2=M_{c}^{'2}-2\mu' \langle \varphi \rangle \,,\,\,~~~M_4^2=M_c^{'2}+2\mu' \langle \varphi \rangle \,.   \label{eq:m3m4_rel}
\end{equation}

\begin{figure}[h!]
\centering
\includegraphics[height=0.4\textwidth]{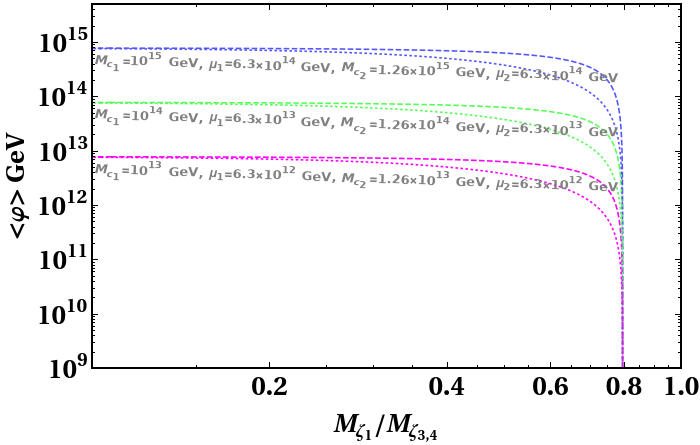}
\caption{\it The variation of mass scales with respect to $\langle\varphi\rangle$ in the plane of ratios $M_{\zeta_1}/M_{\zeta_3}$(dashed) and $M_{\zeta_1}/M_{\zeta_4}(dotted)$.}
\label{fig:m1m34vsphi_param}
\end{figure}

In Fig.\,\ref{fig:m1m34vsphi_param} we have shown
the dependence of the mass ratios of $M_{\zeta_1}/M_{\zeta_3}$ and $M_{\zeta_1}/M_{\zeta_4}$ on $\langle\varphi\rangle$. As we mentioned earlier, the $M_c$ and $M_{c}'$ dictate the mass scale of the $M_{\zeta_{1,2}}$ and $M_{\zeta_{3,4}}$ respectively. The $\mu$ and $\mu'$ are the coupling strengths of $\varphi$ with $\xi$. We fix the parameters $M_c,\,M_{c}',\,\mu,\,\text{and}\, \mu'$ as given in the description of the figure. By fixing these parameters, we choose to vary $\langle \varphi \rangle$ from $10^{9}\, \, \text{to}\,~10^{15}$ GeV. For small values of $\langle\varphi\rangle$, the masses are similarly equal $M_{\zeta_{1}}\simeq M_{\zeta_{2}}$ and $M_{\zeta_{3}}\simeq M_{\zeta_{4}}$, since the mass splitting is small. Consequently the mass ratios $M_{\zeta_1}/M_{\zeta_3}$ and $M_{\zeta_1}/M_{\zeta_4}$ are almost constant until $\langle \varphi \rangle$ approaches the mass scale of $M_c$ and $M_{c}'$. Once the $\langle \varphi \rangle$ approaches the mass scale of $M_c$ and $M_{c}'$ the mass ratios gradually decreases,  since the mass splitting increases. Comparatively, the $M_{\zeta_1}/M_{\zeta_4}$ falls faster than the $M_{\zeta_1}/M_{\zeta_3}$, see Fig.\,\ref{fig:m1m34vsphi_param}\,. The dashed and dotted lines correspond to the mass ratios $M_{\zeta_1}/M_{\zeta_3}$ and $M_{\zeta_1}/M_{\zeta_4}$, respectively.

In presence of both $\xi_{1,2}$ the potential can have mixing terms, {\it i.e.,} $M_{12}^2\, \xi_1^\dagger \xi_2+ \mu_{12} \,\varphi \,\xi_1 \xi_2+H.c.$, which leads to mixing of the four fields $\zeta_i$ after breaking of $B-L$. The diagonalization of the $4\times4$ mixing mass matrix leads to the redefinition of the couplings $\kappa_{\alpha \beta}^i$ with the right-handed neutrinos, let us represent it as $\lambda_{\alpha\beta}^i$. For simplicity we neglect the mixing terms between the $\xi_1$ and $\xi_2$ and work with the four real scalar fields $\zeta_i$ with masses $M_{\zeta_i}$ and complex symmetric Yukawa couplings $\lambda_{\alpha\beta}^i=\lambda_{\beta\alpha}^i$. The corresponding Lagrangian can be re-written as, 
\begin{align}
 \mathcal{L} \  \supset \  \frac{1}{2} \lambda^j_{\alpha \beta}\,\, \zeta_j\, \overline{\nu}_{R,\alpha} \nu^c_{R,\beta} +  \frac{1}{2} \overline{\lambda}^j_{\alpha \beta}\,\, \zeta_j\, \overline{\nu}_{R,\alpha}^c \nu_{R,\beta} \,.
\label{eq:effective_yukawas}
\end{align}

\begin{figure}[H]
	\begin{center}
		\includegraphics[width=0.6\textwidth]{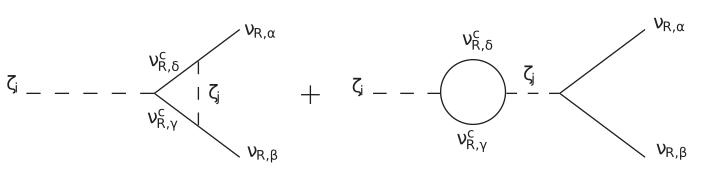}
	\end{center}
		\caption{\it The vertex and self-energy loop diagrams of $\zeta$ decay: $\zeta_i \rightarrow \nu_{R, \alpha} \nu_{R, \beta}$.}
	\label{fig:Feyn_diag}
\end{figure}

In Fig.\,\ref{fig:Feyn_diag} , we show the vertex and self-energy loop diagrams that contribute to the non-zero CP-violation. The decay of the lightest $\zeta_i$ leads to a CP-asymmetry \cite{Heeck:2013vha},
\begin{align}
 \epsilon_i \equiv 2 \, \frac{\Gamma \left(\zeta_i \rightarrow \nu_{R} \nu_{R} \right) - \Gamma \left(\zeta_i \rightarrow \nu_{R}^c \nu_{R}^c \right)}{\Gamma \left(\zeta_i \rightarrow \nu_R \nu_R \right) + \Gamma \left(\zeta_i \rightarrow \nu_R^c \nu_R^c \right)} \,,
\end{align}

The asymmetric contribution from the vertex ($\epsilon_{i}^v$) and self-energy correction ($\epsilon_{i}^s$):
\begin{align}
\begin{split}
 \epsilon^\mathrm{v}_i &= \frac{1}{4\pi} \frac{1}{\text{Tr} (\lambda^{i\dagger} \lambda^i)} \sum_{k\neq i} F (x_k)\, \text{Im} \left[ \text{Tr} \left( \lambda^{i\dagger} \lambda^k \lambda^{i\dagger} \lambda^k \right) \right] ,\\
 \epsilon^\mathrm{s}_i &= -\frac{1}{24\pi} \frac{1}{\text{Tr} (\lambda^{i\dagger} \lambda^i)} \sum_{k\neq i} G (x_k)\, \text{Im} \left[ \left\{ \text{Tr} \left( \lambda^{i\dagger} \lambda^k \right) \right\}^2 \right] ,
\end{split}
\label{eq:asymmetries}
\end{align}

with $x_k \equiv M_{i}^2/M_{k}^2 < 1$ and the functions
\begin{align}
\begin{split}
 F (x) &\equiv \frac{x- \log (1+x)}{x} =  \frac{x}{2} + \mathcal{O} \left( x^2 \right) ,\\
 G (x) &\equiv \frac{x}{1-x} =  x + \mathcal{O} \left( x^2 \right) .
\end{split}
\end{align}

\begin{figure}[h!]
\centering
\includegraphics[height=0.4\textwidth]{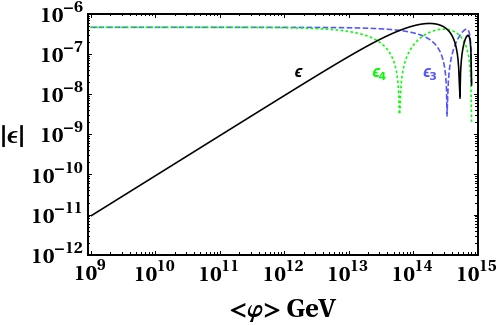}
\caption{\it The dependence of CP-asymmetry with respect to $\langle \varphi \rangle$. The parameter choice are: $M_{c}=1\times 10^{15}$ GeV, $\mu= 6.3\times 10^{14}$ GeV, $M_{c}'=1.26\times 10^{15}$ GeV, $\mu'=7.94\times 10^{14}$ GeV, and  $\text{Tr}[\lambda^{\dagger}\lambda] = 10^{-4}$.}
\label{fig:phi_cpasy}
\end{figure}

The presence of only one scalar $\xi_1$ would lead to vanishing asymmetry $\epsilon^v=0=\epsilon^s$, since $\lambda^{2}=i \lambda^{1}$. For similar reason $\xi_2$ alone does not contribute the CP asymmetry. Thus we require two copies of $\xi$. The total contribution to the total CP-asymmetry come from the lightest $\zeta_1$ with the $\zeta_3$ and $\zeta_4$ mediation, {\it i.e.,} say it, $\epsilon_3$ and $\epsilon_4$, respectively. The non zero contribution come from $\zeta_3$ and $\zeta_4$ and they are opposite in sign, so that $\epsilon^v \propto F(x_3)-F(x_4)$ and $\epsilon^s \propto G(x_3)-G(x_4)$. For the low scales of $\langle \varphi \rangle$, the contributions of $\epsilon_3$ and $\epsilon_4$ are similarly equal in magnitude. They subtract with each other and the resultant total CP-asymmetry $\epsilon$ suppressed, since $\epsilon=\epsilon_3-\epsilon_4$. The dependence of total $\epsilon$ on $\langle \varphi \rangle$ is given in Fig.~\ref{fig:phi_cpasy}\,. The Blue (dashed), Green (dotted), and Black (solid) are $\epsilon_3$, $\epsilon_4$, and $\epsilon$ respectively. Initially, the self-energy contribution $\epsilon_{i}^{s}$ dominates over vertex contribution $\epsilon_{i}^{v}$. When they are equal there is a suppression occurs as shown in the Fig.~\ref{fig:phi_cpasy} for $\epsilon_3$ and $\epsilon_4$. The left and right hand side of the suppression are negative and positive contributions respectively. The subtraction of both magnitudes of $\epsilon_3$ and $\epsilon_4$ gives a negative contribution to the total CP-asymmetry on the left side of the solid Black line suppression. Note that when both the magnitudes of $\epsilon_3$ and $\epsilon_4$ become equal for the first instance (before Black solid line suppression, {\it i.e.,} $\sim 1.5 \times 10^{14}$ GeV), they add up to give maximum magnitude for $\epsilon$, which is still a negative. When both $\epsilon_3$ and $\epsilon_4$ magnitudes become equal for the second instance ($\sim 5 \times 10^{14}$ GeV), they subtract each other and gives a suppression for the $\epsilon$. At the solid Black line suppression point, the CP-asymmetry is zero hence there is no CP-violation. The choice of parameter space is given in the description of the Fig.~\ref{fig:phi_cpasy}\,\footnote{When $2\mu \langle \varphi \rangle$ approach $M_c$ value the $M_{\zeta_1}\rightarrow 0$, that leads to a non-viable scenario.}.

The lightest of $\zeta_i$ decays and creates an asymmetry in the right-handed neutrinos. The abundance of $\zeta_1$ and the generated right-handed neutrino asymmetry can be estimated by solving the following Boltzmann equations,

\begin{eqnarray}
    \frac{d Y_{\zeta_1}}{dz} &=& -\frac{z}{H(M_{\zeta_1})} \Gamma_{\zeta_1}^{\nu_R\nu_R} \left[ Y_{\zeta_1} - Y_{\zeta_1}^{eq} \right], \nonumber\\
    \frac{d Y_{\nu_R}}{dz} &=& \epsilon_1 \frac{z}{H(M_{\zeta_1})} \Gamma_{\zeta_1}^{\nu_R\nu_R} \left[ Y_{\zeta_1} - Y_{\zeta_1}^{eq} \right],
    \label{eq:BEq}
\end{eqnarray}
where $Y_{\nu_R}\equiv n_{\nu_R}/s$ is the total lepton asymmetry, {\it i.e.,} the right-handed neutrino number density $n_{\nu_R}$ relative to the entropy density $s=(2\pi^2/45)g_{*}T^3$. The $Y_{\zeta_1}^{eq}=(45/4\pi^4 g_*)z^2 \mathcal{K}_2(z)$, where $\mathcal{K}_2(z)$ is the modified Bessel function of the second kind. The $\Gamma_{\zeta_1}^{\nu_R\nu_R}$ is the total decay width of the $\zeta_1$ decaying to right-handed neutrinos, 

\begin{equation}
    \Gamma_{\zeta_i}^{\nu_R\nu_R} = \frac{{\rm Tr}(\lambda^{i\dagger} \lambda^i)}{4\pi} M_{\zeta_i}  .
\end{equation}
The $H(M_{\zeta_1})=1.66 \sqrt{g_{*}} \frac{M_{\zeta_1}^2}{M_{\text{Pl}}}$ is Hubble expansion rate at $T=M_{\zeta_1}$.

\begin{figure}[H]
\centering
\includegraphics[height=0.4\textwidth]{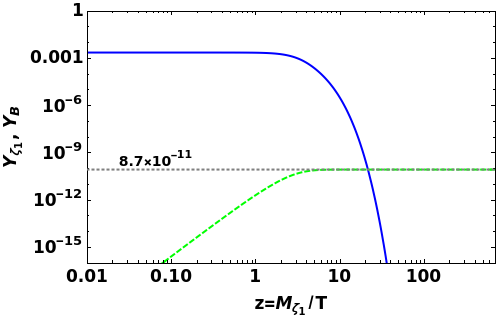}
\caption{\it The abundances of $Y_{\zeta_1}$ (blue) and $Y_{B}$ (green). We consider $M_{c}=8\times 10^{13}$ GeV, $\mu= 10^{13}$ GeV, $M_{c}'=10^{14}$ GeV, $\mu'=10^{13}$ GeV, $\langle \varphi \rangle=10^{12}$ GeV, and  $\text{Tr}[\lambda^{\dagger}\lambda] = 7.2 \times 10^{-4}$. The horizontal dashed line correspond to the observed baryon asymmetry of the Universe.}
\label{fig:yb_z}
\end{figure}

The out-of-equilibrium condition for the decay of the lightest $\zeta_i$ can be estimated as
\begin{equation}
\Gamma(\zeta_i \rightarrow \nu_R \nu_R, \nu_R^c \nu_R^c) \ll H(T\sim M_{\zeta_i}) \simeq  1.66 \sqrt{g_{*}} \frac{M_{\zeta_i}^2}{M_{\text{Pl}}}.
\label{Eq_cond}
\end{equation}
This condition lead to 
\begin{equation}
    \frac{\text{Tr}(\lambda^{i\dagger} \lambda^{i})}{10^{-6}} \ll \frac{M_{\zeta_i}}{10^{11} {\rm GeV}}\,.
\end{equation}
The asymmetry generates in the right-handed neutrinos $Y_{\nu_R}$ after the breaking of $B-L$ and before the electroweak phase transition, so that sphalerons transfer the lepton symmetry to the baryon asymmetry. Initially, the $\zeta_1$ creates the right-handed asymmetry \footnote{We assume that the initial abundance of the right-handed neutrinos to be 0.}, on which the sphalerons do not act. Here we require another term which can transfer the right-handed asymmetry efficiently to left-handed sector. This role played by the term $w_{\alpha\beta} \overline{L}_\alpha \Phi \,\nu_{R,\beta}$, where $\Phi$ is the second Higgs doublet \footnote{We can consider the second Higgs doublet do not acquire a vev, in which case the neutrinos get their mass via SM Higgs $H$. If we consider a small vev to the second Higgs doublet, it can give rise to neutrino mass naturally by softly breaking the new $U(1)$ symmetry via the term $\mu_{12}^2 H^{\dagger} \Phi$ in the potential. This term induces a small vev for $\Phi$, {\it i.e.,} $\langle \Phi \rangle = \mu_{12}^2/M_{\Phi}^2) \langle H \rangle$, which gives rise to small Dirac neutrino masses $M_{\alpha \beta}=\omega_{\alpha \beta} |\langle \Phi \rangle|$.}. This term thermalizes\footnote{If $U(1)_{B-L}$ is a gauge symmetry, the $Z'$ interactions (and $\Phi$) can thermalize the $\nu_R$. If $U(1)_{B-L}$ is a global symmetry, the $\Phi$ can thermalize the $\nu_R$.} the $\nu_R$ and transfer the asymmetry from right-handed sector to left-handed sector, {\it i.e.,} $\Delta_{\nu_R} \rightarrow \Delta_L$. The second scalar doublet couples with neutrino sector due to an additional $U(1)$ symmetry under which only $\Phi$ and $\nu_R$ are charged. We solve the Boltzmann equations, given in Eq.\,\ref{eq:BEq}\,, to estimate the abundance of $Y_{\zeta_1}$ and the baryon asymmetry $Y_B$, shown in Fig.\,\ref{fig:yb_z}\,. The total lepton asymmetry can be expressed as 
\begin{equation}
Y_{\nu_R}\equiv \frac{n_{\nu_R}}{s} = \epsilon_i Y_{\zeta_1}^{eq}\sim \frac{\epsilon_{i}^v+\epsilon_i^s}{g_{*}}.    
\end{equation}
We assume equilibrium of right-handed neutrino with the SM particles as well as sphalerons, we use the chemical potentials to describe the plasma including the chemical potential for the right-handed neutrinos, resulting in the equilibrium condition $3 B+L=0$ or 
\begin{equation}
    Y_B=\frac{1}{4} Y_{B-L},~~~~~~ Y_L=-\frac{3}{4} Y_{B-L},
\end{equation}
for three generations (and an arbitrary number of Higgs doublets). Once the right-handed neutrino asymmetry produced a (1/4) part of it transfer to the baryon asymmetry of the Universe. The orange (solid) and blue (dashed) lines are the abundance of the $\zeta_1$ and the baryon asymmetry $Y_B$ respectively.

\medskip

\section{Primordial GW spectrum from Local Cosmic Strings} \label{sec:GW_local_cs}

Cosmic Strings (CS) may originate as fundamental objects such as in String Theory \cite{Copeland:2009ga,Polchinski:2004ia} as well as topological defects in field theory during a symmetry-breaking phase transition. We will focus on cosmic strings that were formed in the early Universe due to the spontaneous breaking of a local $U(1)$ symmetry, this is sometimes called field theory strings. The simplest model where cosmic strings cmay arise is a $U(1)$ local gauge or global theory in the context of a complex scalar field denoted by $\varphi$. The Lagrangian density for such a theory can be written as, 
\begin{equation}
    \mathcal{L}=D_\mu\varphi^*D^\mu\varphi-\frac{1}{4}F_{\mu\nu}F^{\mu\nu}-V(\varphi)
\end{equation}
where $D_\mu\varphi=\partial_\mu\varphi+\mathrm{i}eA_\mu\varphi $ is the covariant derivative and $F_{\mu\nu}=\partial_\mu A_\nu-\partial_\nu A_\mu $ is the field tensor. 

$V(\varphi)$ is the typical Mexican hat potential written as,
\begin{equation}
    V(\varphi)=\frac{1}{2}\lambda'(\varphi^*\varphi-\frac{1}{2}\eta^2)^2
\end{equation}
where $\lambda'$ is the self-quartic coupling and $\langle \varphi \rangle = \eta$ is the vacuum expectation value (VEV) of the scalar field. 

\begin{figure}[H]
    \centering
    \includegraphics[width=0.7\linewidth]{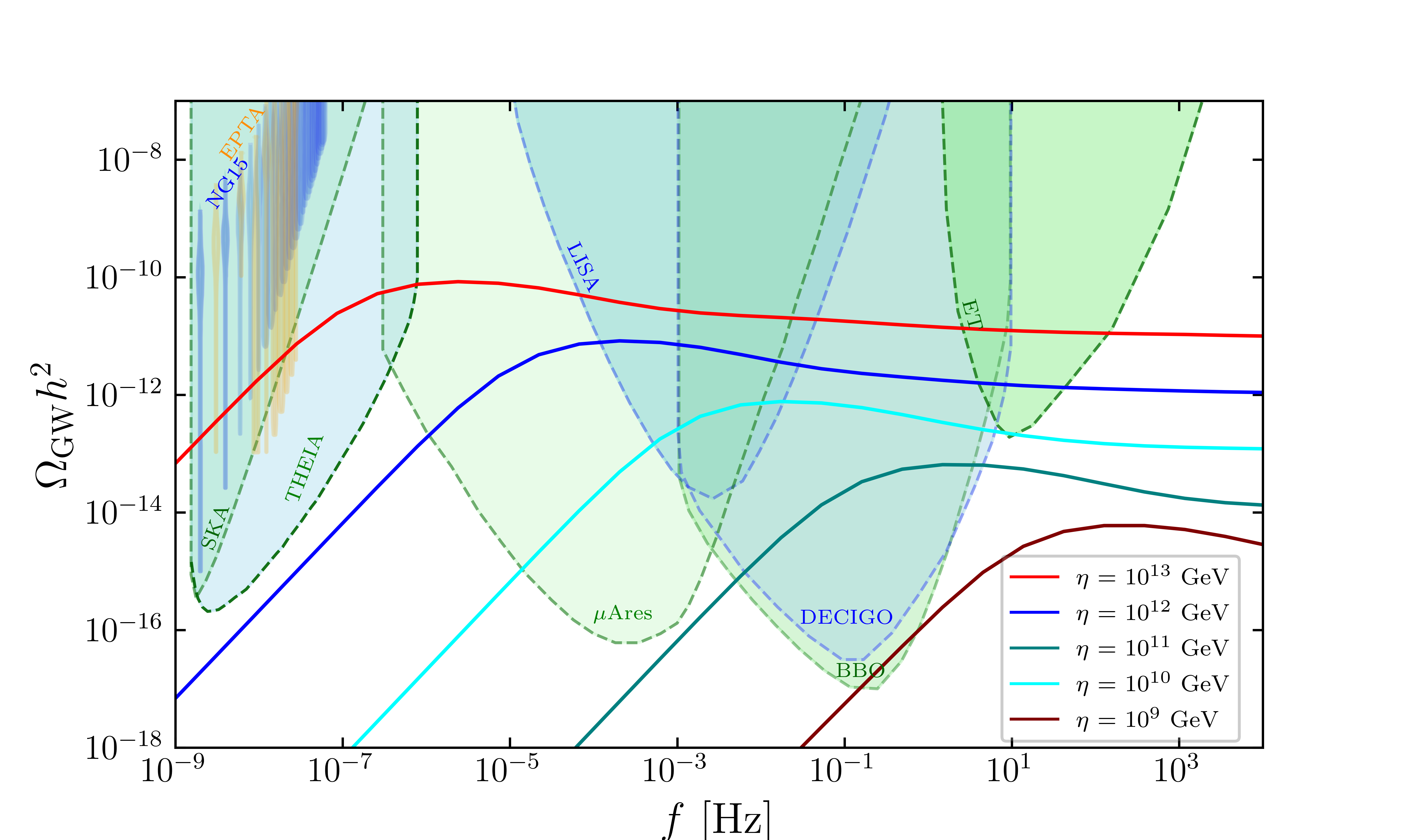}
    \caption{\it Variation of the SGWB spectra from local cosmic strings with respect to the vacuum expectation value (VEV) $\langle \varphi \rangle = \eta$.}
    \label{fig:GW}
\end{figure}
Cosmic strings are essentially field configurations at the top of this Mexican hat potential. The CS have a negligible thickness (with widths given by approximately $1/(\sqrt{\lambda}\eta$)) and hence their dynamics in an expanding Universe are well described by something known as the Nambu-Goto action, in its zero-width approximation. The mass per unit length $\mu$ of local CS is given by \cite{Gouttenoire:2019kij}, 
\begin{equation}
    \mu=2\pi n \eta^2
\end{equation}
where $\eta$ is the vacuum expectation value of the scalar field constituting the CS and n is the (integer-valued) winding number of the cosmic string formed. Only $n=1$ is considered to be a stable configuration \cite{Bettencourt:1996qe,Ghoshal:2023sfa}. 
The local CS network starts forming at a temperature, 
\begin{equation}\label{local CS network form}
    T_{\mathrm{CS}}\simeq\eta\simeq10^{11} \mathrm{GeV}\left(\frac{G\mu}{10^{-15}}\right)^{1/2}
\end{equation}
which is also the temperature of a $U(1)$-breaking phase transition. $G$ is Newton's constant and the dimensionless parameter $G\mu$ is knwon as the string tension.

Straight infinitely long strings are stable against decay due to their topological nature, therefore they contribute negligibly towards the GW spectra compared to cosmic string loops. Local CS loops, formed due to intercommutation events of CS, are free of any topological charge, so they contribute significantly towards GW radiation emission \cite{Vilenkin:2000jqa}. The GW radiation power for CS loops is given by,
\begin{equation}
    P_{\mathrm{GW}}=\Gamma G\mu^2
\end{equation}
where $\Gamma$ is the total GW emission efficiency of loops and is determined typically from the Nambu-Goto simulation. This has been estimated to be $\Gamma\simeq70$ \cite{Blanco-Pillado:2013qja}.

The GW spectrum from Local CS Loops observed today is given follows \cite{Gouttenoire:2019kij}:
\begin{equation}
    \Omega_{\mathrm{GW}}(f) \equiv \frac{f}{\rho_c}\left|\frac{d \rho_{\mathrm{GW}}}{d f}\right|=\sum_k \Omega_{\mathrm{GW}}^{(k)}(f)
\end{equation}

where

\begin{equation}\label{Local GW1}
    \begin{aligned}\Omega_{\mathrm{GW}}(f)=\sum_{k}\frac{1}{\rho_{c}}\int_{t_{\mathrm{osc}}}^{t_{0}}d\tilde{t}\int_{0}^{1}d\alpha\cdot \Theta\left[t_{i}-\frac{l_{*}}{\alpha}\right]\cdot\Theta[t_{i}-t_{\mathrm{osc}}]\cdot\left[\frac{a(\tilde{t})}{a(t_{0})}\right]^{4}\cdot P_{\mathrm{GW}}^{(k)}\times\\\times\left[\frac{a(t_{i})}{a(\tilde{t})}\right]^{3}\cdot\mathcal{P}_{\mathrm{loop}}(\alpha)\cdot\frac{dt_{i}}{df}\cdot\frac{dn_{\mathrm{loop}}}{dt_{i}}\end{aligned}
\end{equation}

The first Heaviside function $\Theta\left[t_{i}-\frac{l_{*}}{\alpha}\right]$ ensures that loops smaller than critical length $l_{*}$ won't contribute towards the GW spectrum as their main decay channel is through massive particle production. $l_* \equiv l_c$ is used for cusps and $l_k$ is used for kinks.

\begin{equation}
    l_c \equiv \beta_c \: \frac{\mu^{-1/2}}{(\Gamma G \mu)^2}; \quad
    l_k \equiv \beta_k \: \frac{\mu^{-1/2}}{\Gamma G \mu}
    \label{eq:cusp_kink_lengths}
\end{equation}

$\beta_c$ and $\beta_k$ are some $\mathcal{O}(1)$ numbers.  The second Heaviside function $\Theta[t_{i}-t_{\mathrm{osc}}]$ get rid of all those loops that were formed before the network formation (Eq. \ref{local CS network form}) or which form during the friction dominated era, 
\begin{equation}
    T \lesssim T_{\mathrm{fric}}=\frac{4\times10^9 \mathrm{GeV}}{\beta}\left(\frac{g_*}{100}\right)^{1/2}\left(\frac{G\mu}{10^{-11}}\right)
\end{equation}
and $t_{\mathrm{osc}}=\operatorname{Max}\left[t_{fric}, t_F\right]$, where $t_F$ is the time of CS network formation, defined as $\sqrt{\rho_{tot}\left(t_F\right)} \equiv \mu$ where $\rho_{tot}$ is the total energy budget of the Universe. In the presence of friction, the string motion is damped at high temperature until the time $t_{fric}$.
$dn_{\mathrm{loop}}/dt_i$ expresses the rate of formation of loops with a distribution size $\mathcal{P}_{\mathrm{loop}}(\alpha)$. These objects initially redshift as $a^{-3}$ before radiating GW with power $P_{\mathrm{GW}}^{(k)}$ after which they dilute as $a^{-4}$. Using such simplifications, one may express Eq. \ref{Local GW1} as \cite{Gouttenoire:2019kij}:

\begin{equation}\label{Local GW2}
    \Omega_{\mathrm{GW}}^{(k)}(f)=\frac{1}{\rho_{c}}\cdot\frac{2k}{f}\cdot\frac{\mathcal{F}_{\alpha}}{\alpha(\alpha+\Gamma G\mu)}\int_{t_{\mathrm{osc}}}^{t_{0}}d\tilde{t} \frac{C_{\mathrm{eff}}(t_{i})}{t_{i}^{4}}\left[\frac{a(\tilde{t})}{a(t_{0})}\right]^{5}\left[\frac{a(t_{i})}{a(\tilde{t})}\right]^{3}\Theta(t_{i}-t_{\mathrm{osc}})\Theta(t_{i}-\frac{l_{*}}{\alpha})
\end{equation}

$\alpha ()$ is the loop length at the time of formation and $\mathcal{F}_{\alpha}$ gives the fraction of loops that forms with size $\alpha$. $t_i$ is the time of loop production which is dependent on the emission time $\tilde{t}$ leading to the observed frequency today to be, 
\begin{equation}
    t_i(f,\tilde{t})=\frac{1}{\alpha+\Gamma G\mu}\left[\frac{2k}{f}\frac{a(\tilde{t})}{a(t_0)}+\Gamma G\mu \tilde{t}\right]
\end{equation}
$t_0$ is the present time. $C_{eff}(t_i)$ is the loop-production efficiency given by,
\begin{equation}
    \tilde C_{\mathrm{eff}}\equiv\sqrt{2} C_{\mathrm{eff}}(t)=\frac{\tilde{c} \bar{v}(t)}{\xi^3(t)}
\end{equation}
where $\tilde{c}=0.23\pm0.04$ is a phenomenological parameter which quantifies the loop chopping efficiency factor. $\bar{v}$ is the root mean square speed of the string loops and $\xi$ is defined as the quantity $\xi\equiv L/t$, where $L$ is the correlation length of the cosmic string.

Figure \ref{fig:GW} shows the full GW spectrum due to local CS computed using Eq. \ref{Local GW2} assuming a $\Lambda$CDM Universe assuming the strings were formed after the end of primordial cosmic inflation. The GW spectrum can be broadly classified into a flat high-frequency regime and the radiation-to-matter transition at the low-frequency regime. Higher frequencies correspond to earlier and earlier in times i.e., it corresponds to a radiation-dominated era. The local CS loops that were produced and started GW emission during the radiation-dominated era have a nearly flat GW spectrum with amplitude typically given by,
\begin{equation}
    \Omega_{\mathrm{std}}^{\mathrm{CS}}h^2\simeq\Omega_rh^2\mathcal{G}(\tilde{T}_{\mathrm{M}})\left(\frac{\eta}{M_{\mathrm{pl}}}\right)
\end{equation}

where $\Omega_r h^2 \simeq 4.2 \times 10^{-5}$ \cite{ParticleDataGroup:2020ssz}. $\tilde{T_M}$ is the temperature of the Universe at the time of the maximal GW emission $\tilde{t_M}$,
\begin{equation}
    \tilde{t_M}=\frac{\alpha}{2\Gamma G\mu} t_i \,.
\end{equation}

Interestingly, the deviation from the nearly flat spectrum occurs due to a change in the number of relativistic degrees of freedom which is approxmated by,

\begin{equation}
\mathcal{G}(T) \equiv\left(\frac{g_*(T)}{g_*\left(T_0\right)}\right)\left(\frac{g_* s\left(T_0\right)}{g_{* s}(T)}\right)^{4 / 3}=0.39\left(\frac{106.75}{g_*(T)}\right)^{1 / 3}\,.
\end{equation}


\section{Numerical results}\label{sec:Num_res}

All kinds of Gravitational Wave experimental efforts can be sorted into:

\begin{enumerate}
    \item \textbf{Ground based interferometers:} \textit{Laser Interferometer Gravitational-wave Observatory} (LIGO)~\cite{LIGOScientific:2016aoc,LIGOScientific:2016sjg,LIGOScientific:2017bnn,LIGOScientific:2017vox,LIGOScientific:2017ycc,LIGOScientific:2017vwq}, \textit{Advanced} LIGO (a-LIGO)~\cite{LIGOScientific:2014pky,LIGOScientific:2019lzm},  \textit{Einstein Telescope} (ET)~\cite{Punturo_2010,Hild:2010id}, \textit{Cosmic Explorer} (CE)~\cite{Reitze:2019iox}.
    \item \textbf{Space based interferometers: }$\mu$-ARES~\cite{Sesana:2019vho}, \textit{Laser Interferometer Space Antenna} (LISA)~\cite{amaroseoane2017laser,Baker:2019nia}, \textit{Big-Bang Observer} (BBO)~\cite{Corbin:2005ny,Harry_2006}, \textit{Deci-Hertz Interferometer Gravitaitonal-wave Observatory} (DECIGO)~\cite{Yagi:2011yu}, \textit{Upgraded} DECIGO (U-DECIGO)~\cite{Seto:2001qf,Kawamura_2006,Yagi:2011wg}.
    \item \textbf{Pulsar Timing Arrays (PTA):} \textit{European Pulsar Timing Array} These are recast of star surveys (EPTA)~\cite{Kramer:2013kea,Lentati:2015qwp,Babak:2015lua}, \textit{Square Kilometre Array} (SKA)~\cite{Janssen:2014dka,Weltman:2018zrl,Carilli:2004nx}, \textit{North American Nanohertz Observatory for Gravitational Waves} (NANOGrav)~\cite{McLaughlin:2013ira,NANOGRAV:2018hou,Aggarwal:2018mgp,Brazier:2019mmu,NANOGrav:2020bcs}.
\end{enumerate} 

\subsection{Gravitational Wave as probe of Dirac Leptogenesis}

\begin{figure}[b]
\centering
\includegraphics[height=0.55\textwidth]{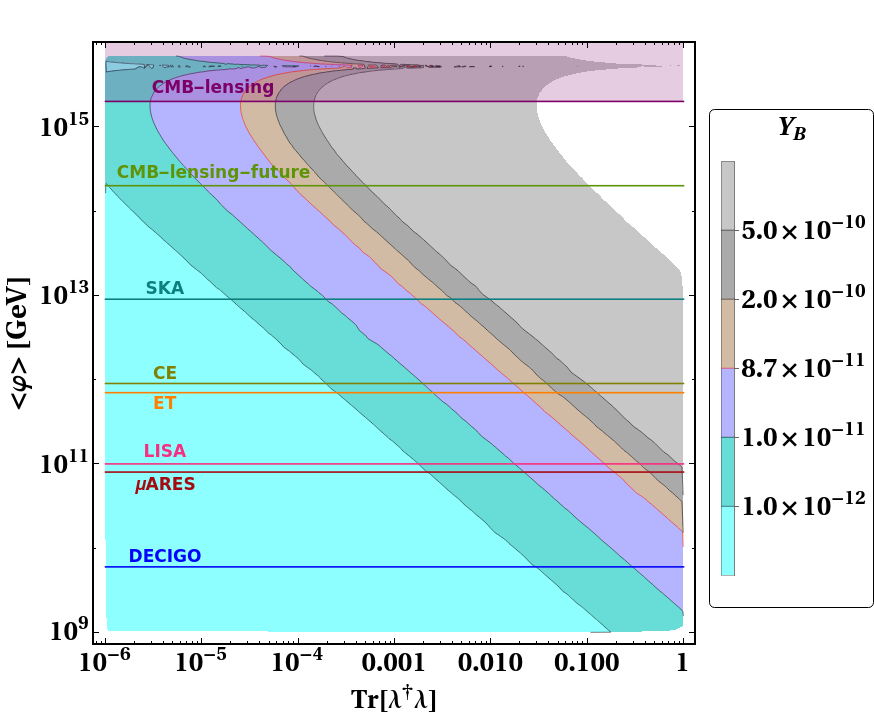}
\caption{\it \textbf{Dirac Leptogenesis}:  Parameter space probed by individual GW detectors in the plane of $\text{Tr}[\lambda^{\dagger}\lambda]$ and $\langle \varphi \rangle$. We fix the parameters: $M_{c}=10^{16}$ GeV, $\mu=6.31\times 10^{15}$ GeV, $M_{c}'=1.26\times 10^{16}$ GeV and $\mu'=7.94\times 10^{15}$ GeV.}
\label{fig:lambda_vs_phi}
\end{figure}

The vev of the scalar $\langle\varphi\rangle$ plays a significant role in Dirac Leptogenesis and the Gravitational waves. Another important parameter that plays key role in Dirac Leptogenesis is $\text{Tr}[\lambda^{\dagger}\lambda]$, which quantifies the amount of CP-violation. To study the allowed parameter space of the baryon asymmetry we plot Fig.\,\ref{fig:lambda_vs_phi}\,, where the contours of baryon asymmetry are drawn in the plane of $\text{Tr}[\lambda^{\dagger}\lambda]$ and $\langle\varphi\rangle$. We see the observed baryon asymmetry of the Universe is allowed for a wide range of $\text{Tr}[\lambda^{\dagger}\lambda]$ and $\langle \varphi \rangle$. Here we fix the parameter as: $M_{c}=10^{16}$ GeV, $\mu=6.31\times 10^{15}$ GeV, $M_{c}'=1.26\times 10^{16}$ GeV and $\mu'=7.94\times 10^{15}$ GeV. As the CP-asymmetry parameter $\epsilon$ is proportional to $\text{Tr}[\lambda^{\dagger}\lambda]$, for a particular contour of baryon asymmetry if we decrease $\text{Tr}[\lambda^{\dagger}\lambda]$ the resultant baryon asymmetry decreases. To compensate that the $\langle \varphi \rangle$ has to increase so that the $\epsilon$ increases, as discussed in Fig.\,\ref{fig:phi_cpasy}\,, which results in desired contour of baryon asymmetry as shown in the figure. For the larger values of $\text{Tr}[\lambda^{\dagger}\lambda]$, the $\langle \varphi \rangle$ takes lower values for the same reason. Note that for higher values of $\langle \varphi \rangle$ the contour of baryon asymmetry move towards the larger values of $\text{Tr}[\lambda^{\dagger}\lambda]$ due to a suppression occurs in the total CP-asymmetry, as discussed in Fig.~\ref{fig:phi_cpasy}\,. Hence to compensate that $\text{Tr}[\lambda^{\dagger}\lambda]$ increases. At the suppression point the CP-asymmetry is zero, hence no baryon asymmetry. The white patch region shows the excessive amount of baryon asymmetry.  For a decrease of the mass scales described in the caption of the figure\footnote{The choice of high mass scales in the Fig.\,\ref{fig:lambda_vs_phi} is to consider the scale of $\langle \phi \rangle$ to cover observations involving CMB-lensing, CMB-lensing-future, etc,.} by an order the contours shift down by an order while leaving $\text{Tr}[\lambda^{\dagger}\lambda]$ intact. 

In Fig.\,\ref{fig:lambda_vs_phi} we also show various GW detectors sensitivity limits come from DECIGO, $\mu$ARES, LISA, ET, CE, SKA, etc,. The region above the particular sensitivity limit can be probed by the corresponding GW detectors with \textit{Signal-to-Noise Ratio} (SNR) estimations as shown in Appendix \ref{SNR}\,. We observe that the Dirac Leptogenesis can be probed by the most of GW detectors like DECIGO, $\mu$ARES, LISA, ET, CE, SKA and CMB-lensing-future. The region above CMB-lensing is ruled out.

\subsection{Gravitational Wave as probe of Majorana Leptogenesis}

\begin{figure}[h!]
\centering
\includegraphics[height=0.55\textwidth]{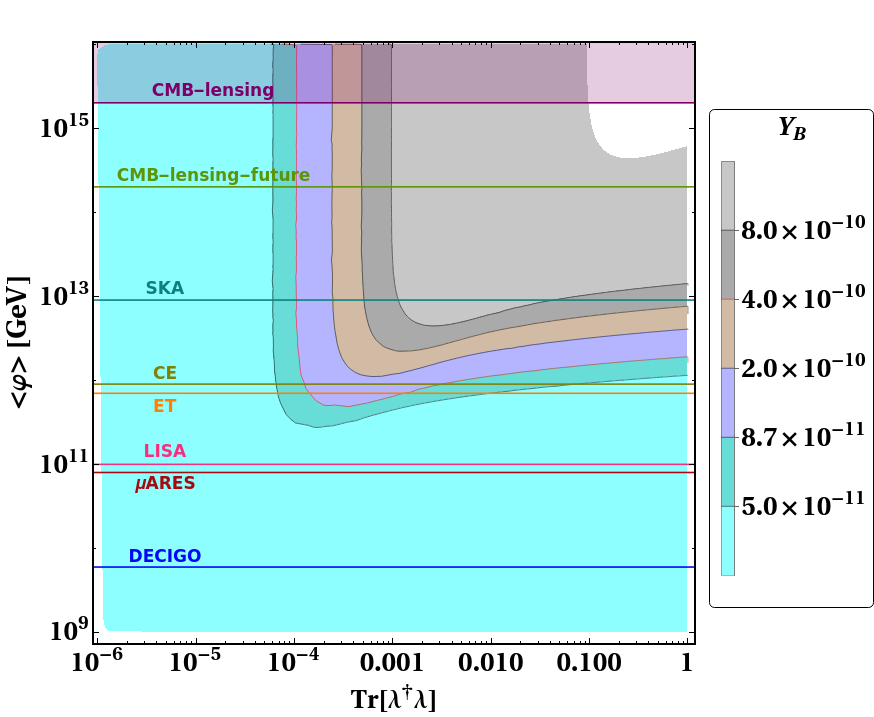}
\caption{\it  \textbf{Majorana Leptogenesis}: Parameter space probed by individual GW detectors in the plane of $\text{Tr}[\lambda^{\dagger}\lambda]$ and $\langle \varphi \rangle$. We fix the parameters: $y_m=1$, $M_{N_2}/M_{N_1}=10^2$.}
\label{fig:maj_lambda_vs_phi}
\end{figure}

In Type-I seesaw mechanism, the heavy right-handed neutrinos $N_i,\,\{i=1, 2, 3\}$ introduced to address the smallness of the light neutrino masses. The relevant Lagrangian given as follows,
\begin{equation}
    \mathcal{L} \supset y_{i\alpha} \overline{N_i} H^{\dagger} L_{\alpha}+\frac{1}{2} M_N \overline{N}N+h.c.,
\end{equation}
where $y$ is the Yukawa coupling and $M_N=y_m \langle \varphi \rangle$ is a diagonal mass matrix of the right-handed neutrinos. In standard Majorana Leptogenesis the heavy right-handed neutrinos couples with $L$ and $H$. The lightest of heavy right-handed neutrinos decays $N_1\rightarrow L + H$, while fulfilling the Sakharov conditions, and generates lepton asymmetry. A part of the lepton asymmetry transfer to the baryon asymmetry by sphaleron transitions. The Boltzmann equations governing the generation of the lepton asymmetry can be given as follows,
\begin{eqnarray}
    \frac{d Y_{N_1}}{dz} & = & -D(z) \,(Y_{N_1}-Y_{N_1}^{eq}), \\
    \frac{d Y_{B-L}}{dz} & = & - \epsilon_1 \, D(z)\, (Y_{N_1}-Y_{N_1}^{eq})-W_{ID}(z) Y_{B-L}\,,
\end{eqnarray}
where 
\begin{eqnarray}
    z & = & M_1/T\,,\\
    D(z) & = & K_1 z \frac{\mathcal{K}_1(z)}{\mathcal{K}_2 (z)}\,,\\
    W_{ID} (z) & = & \frac{1}{4} K_1 z^3 \mathcal{K}_1(z)\,,\\
    K_1 & = & \frac{\Gamma_{N_1}}{H(M_{N_1})}.
\end{eqnarray}
where $\mathcal{K}_{1,2}$ are modified Bessel functions of first and second kind. The equilibrium number density defined as 
\begin{eqnarray}
    Y_{N_1}^{eq}=\frac{3}{4}\zeta(3) \frac{45}{2\pi^4 g_*}z^2 \mathcal{K}_2(z),
\end{eqnarray}
where $g_*$ is the relativistic degrees of freedom.
For a hierarchical right-handed neutrino masses the amount of CP-violation is calculated by the CP-asymmetry parameter $\epsilon_1$ as,
\begin{eqnarray}
        \epsilon_1= \frac{3}{16 \pi (y^{\dagger}y)_{ii}} \sum_{j\neq i} \text{Im}\left[(y^{\dagger}y)_{ji}^{2}\right] \left( \frac{M_{N_i}}{M_{N_j}} \right).
        \label{epsilon1}    
\end{eqnarray}
Once the lepton asymmetry has generated, part of it can be transfer to the baryon asymmetry,
 \begin{equation}
     Y_{B}=-\frac{28}{79} Y_{B-L}\,\,.
 \end{equation}
We examine the Majorana Leptogenesis in Fig.\,\ref{fig:maj_lambda_vs_phi}\,, the baryon asymmetry portrayed as contours in the plane of ${\rm Tr}[\lambda^\dagger \lambda]$ and $\langle\varphi \rangle$ while imposing various GW detector constraints\footnote{ We consider the scale of B-L breaking, $\langle\varphi \rangle \gtrsim10^{9}$ GeV, to make $M_N$ scale consistent with Davidson-Ibarra bound \cite{Davidson:2002qv}. }. For large values of ${\rm Tr}[\lambda^\dagger \lambda]$ and $\langle\varphi \rangle$, we observe the correct baryon asymmetry of the Universe. At higher scales of $\langle\varphi \rangle$ the contours of the baryon asymmetry behaves as independent of ${\rm Tr}[\lambda^\dagger \lambda]$. But as we move towards lower $\langle\varphi \rangle$ the washouts dominates and the baryon asymmetry decreases. To achieve the desired baryon asymmetry contour at low $\langle \varphi \rangle$ the coupling ${\rm Tr}[\lambda^\dagger \lambda]$ has to increase, therefore contour lines moves towards right. Meanwhile for large values of ${\rm Tr}[\lambda^\dagger \lambda]$ the washouts increases too and to compensate that the $\langle \varphi \rangle$ has to increase hence the contour line rises slightly at higher ${\rm Tr}[\lambda^\dagger \lambda]$. The parameter choice is given in the description of Fig.\ref{fig:maj_lambda_vs_phi}\,. The contours of BAU shift down to low scales of $\langle\varphi\rangle$ and $\text{Tr}[\lambda^{\dagger}\lambda]$ upon decreasing the mass ratio of RHNs, $M_{N_2}/M_{N_1}$. In Fig.\,\ref{fig:maj_lambda_vs_phi} we show various GW detectors sensitivity limits come from DECIGO, $\mu$ARES, LISA, ET, CE, SKA, etc,. The region above the particular sensitivity limit can be probed by the corresponding GW detectors with \textit{Signal-to-Noise Ratio} (SNR) estimations as shown in Appendix \ref{SNR}\,.

\subsection{Comparison: Dirac versus Majorana Leptogenesis}

In general, the $B-L$ number is violated in Majorana Leptogenesis, whereas in Dirac Leptogenesis the $B-L$ is conserved\,\cite{Dick:1999je, Narendra:2017uxl}. In current scenario, the Dirac Leptogenesis the $B-L$ number violates similar to the Majorana scenario\,\cite{Heeck:2013vha}. The origin of lepton asymmetry is different in both scenarios. In Dirac Leptogenesis case a heavy scalar generates asymmetry in right-handed neutrinos and then transfer to the lepton asymmetry, whereas in Majorana Leptogenesis case a heavy right-handed neutrino generates the lepton asymmetry. In the following, to establish a comparison between these scenarios we compare the mass scales of the Leptogenesis.

\begin{figure}[h!]
\centering
\includegraphics[height=0.55\textwidth]{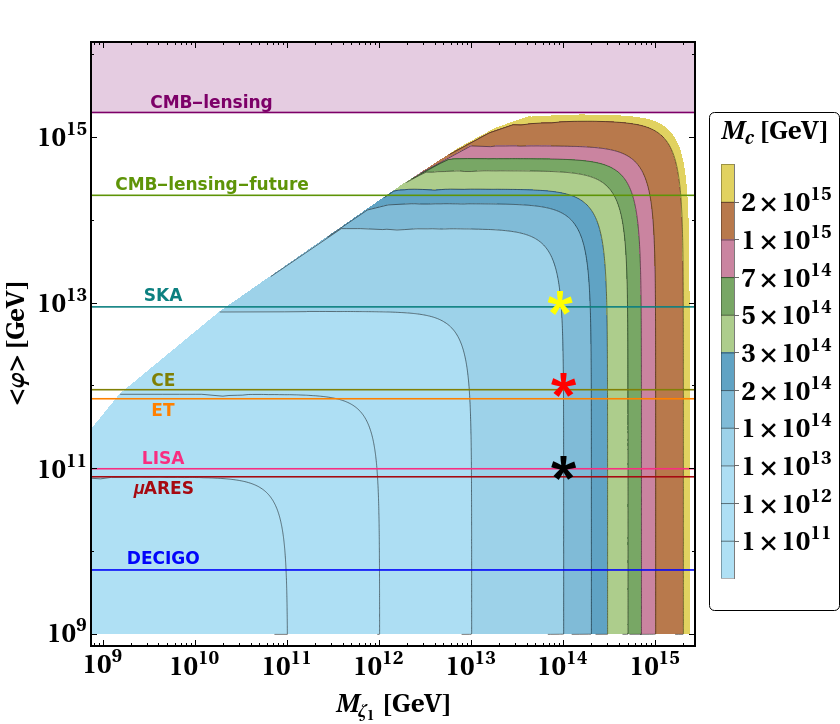}
\caption{\it \textbf{Dirac Leptogenesis:} Parameter space probed by individual GW detectors in the plane of $M_{\zeta_1}$ and $\langle \varphi \rangle$. We fix the parameters: $\mu=0.63 \times M_c$ GeV, $M_{c}'=1.26 \times M_{c} $ and $\mu'=0.63 \times M_{c}'$ GeV. The Black, Red and Magenta star points represent the BPs provided in Table \ref{table_Dirac_BPs}. The plot represents the correlation between the scale of Leptogenesis and the scale of seesaw and satisfies Eq.\,\ref{eq:m1m2_rel}\,.}
\label{fig:mzeta1_vs_phi}
\end{figure}

\begin{table}[H]
\begin{center}
\begin{tabular}{|c|c|c|c|c|c|c|}
\hline
Parameters & $M_{\zeta_1}$ [GeV] & $M_c$ [GeV] &  $\mu$ [GeV] & $\langle\varphi\rangle$ [GeV] & $\text{Tr}[\lambda^{\dagger} \lambda]$ & $Y_B$ \\
\hline
BP 1 (\textcolor{yellow}{$\star$}) & $9.3\times10^{13}$  &  $10^{14}$  &  $6.3\times10^{13} $ & $10^{13}$ &  $2.8 \times 10^{-5}$ & $8.7\times 10^{-11}$ \\
\hline
BP 2 (\textcolor{red}{$\star$}) & $9.9\times10^{13}$ &  $10^{14} $  &  $6.3\times10^{13} $ & $10^{12}$ &  $1.6 \times 10^{-4}$ & $8.7\times 10^{-11}$  \\
\hline
BP 3 (\textcolor{black}{$\star$}) & $9.9\times10^{13}$ &  $10^{14}$  &  $6.3\times10^{13}$ & $10^{11}$ &  $1.5 \times 10^{-3}$ & $8.7\times 10^{-11}$  \\
\hline
\end{tabular}
\end{center}
\caption{\it \textbf{Dirac Leptogenesis:} Benchmark Points (BPs) are corresponding to the star($*$) symbol points Black (BP 1), Red (BP 2), and Yellow (BP 3) represented in Fig. \ref{fig:mzeta1_vs_phi}. }
\label{table_Dirac_BPs}
\end{table}

The Fig.\,\ref{fig:mzeta1_vs_phi}\,, depicts the mass the scale of Dirac Leptogenesis $M_{\zeta_1}$ with respect to $\langle \varphi \rangle$ with the contours of various choices of $M_c$. Note that mass scale of $M_{\zeta_1}$ depends on the choice of $M_c$. For a fixed value of $M_c$, the $M_{\zeta_{1}}$ is barely constant with the increment of $\langle \varphi \rangle$, but as $\langle \varphi \rangle$ approach $M_c$ the $M_{\zeta_{1}}$ gradually decreases, see Eq.\,\ref{eq:m1m2_rel}\,. When the $2\mu \langle \varphi \rangle$ similar to the $M_c^2$ the $M_{\zeta_1}$ drops drastically and can approach zero with high fine-tuning\footnote{We use the fine-tuning of $2\mu\langle\varphi\rangle$ with $M_c^2$ of the order $\sim10^{-5}$ to get such low $M_{\zeta_1}$ shown in Fig.\,\ref{fig:mzeta1_vs_phi}\,.}. For lower values of $M_c$ the allowed mass scale for $M_{\zeta_1}$ decreases. The white patch region on the top left is redundant. The $\mu$ is taken one order lesser than $M_c$, {\it i.e., $\mu=0.63\times M_c$} as an arbitrary choice. 

We provide three Benchmark Points in Table.\,\ref{table_Dirac_BPs} that are well agreement with the GWs originated from the cosmic strings and the Dirac Leptogenesis. We consider three representative points for the $\langle \varphi \rangle=\{10^{13} (\text{Yellow}),\, 10^{12} (\text{Red}),\,10^{11}  (\text{Black})\}$ GeV, from Fig.\,\ref{fig:GW}, the GW signal which overlap with the LISA and ET sensitivity. Based on $\langle \varphi \rangle$ we consider three representative points for the $M_c$ that give different mass scales for the $M_{\zeta_1}$. The other parameter choices are taken as specified in the description of the Fig.\,\ref{fig:mzeta1_vs_phi}\,. We observe that three BPs well agree with the observed BAU. These BPs fall near the sensitivity of SKA, CE, ET, LISA and $\mu ARES$ future detection region. The BPs can be detectable by SKA, CE, ET, LISA, $\mu ARES$ and DECIGO SNR threshold sensitivities and can give a hint towards the Dirac nature of light neutrinos and Leptogenesis. 

\begin{figure}[h!]
\centering
\includegraphics[height=0.55\textwidth]{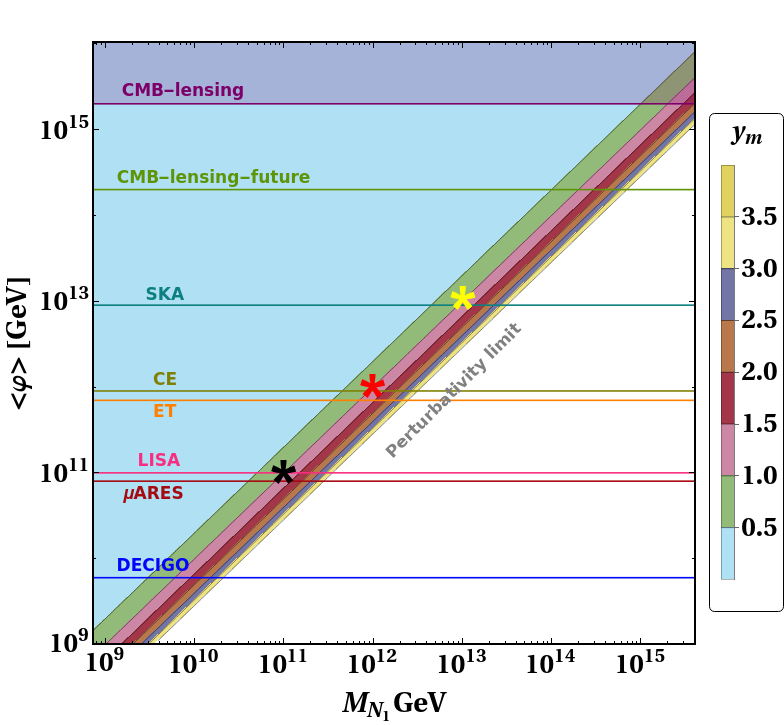}
\caption{\it \textbf{Majorana Leptogenesis:} Parameter space probed by individual GW detectors in the plane of $M_{N_1}$ and $\langle \varphi \rangle$. The mass ratio: $M_{N_2}/M_{N_1}=10^2$. The Black, Red and Yellow star points represent the BPs provided in Table \,\ref{table_Maj_BPs}. The plot represents the correlation between the scale of leptogenesis and the scale of seesaw.}
\label{fig:maj_mn_vs_phi}
\end{figure}

We also provide three BPs for Majorana Leptogenesis for a comparison with the Dirac Leptogenesis parameter space. Here also we consider three representative points for the $\langle \varphi \rangle=\{10^{13}(\text{Yellow}),\, 10^{12}(\text{Red}),\,10^{11}(\text{Black})\}$ GeV, from the Fig.\,\ref{fig:GW}, the GW signal which overlap with the LISA and ET. For comparison we consider the mass scale of the right handed neutrino $M_{N_1}$ is similar to the mass scale of Dirac Leptogenesis, {\it i.e.,} $M_{\zeta_1}$. In Fig.\ref{fig:maj_mn_vs_phi}, we show the mass scale of the Majorana Leptogenesis with respect to $\varphi$ with the contours of Yukawa coupling $y_m$. The contours are straight lines, since the $M_{N_1}$ and $\langle\varphi\rangle$ are proportional. We observe that two BPs with $\langle\varphi \rangle=10^{13},\,10^{12}$ GeV well agree with the observed BAU, for the mass hierarchy $M_{N_2}/M_{N_1}=10^{2}$, whereas the $\langle \varphi \rangle=10^{11}$ GeV gives insufficient baryon asymmetry. The two BPs fall near the sensitivity of SKA, CE and ET future detection region. These BPs can detectable by the SKA, CE, ET, LISA, $\mu ARES$ and DECIGO SNR threshold sensitivities and can give a hint towards the Dirac nature of light neutrinos and Leptogenesis.

\begin{table}[H]
\begin{center}
\begin{tabular}{|c|c|c|c|c|c|}
\hline
Parameters & $M_{N_1}$ [GeV] & $y_m$ &  $\langle\varphi\rangle$ [GeV] & $\text{Tr}[\lambda^{\dagger} \lambda]$ & $Y_B$ \\
\hline
BP 1 (\textcolor{yellow}{$\star$}) & $10^{13}$  &  1  & $10^{13}$ &  $1\times 10^{-4}$ & $8.7\times 10^{-11}$ \\
\hline
BP 2 (\textcolor{red}{$\star$}) & $10^{12}$ &  1  & $10^{12}$ &  $1.2\times 10^{-4}$ & $8.7\times 10^{-11}$  \\
\hline
BP 3 (\textcolor{black}{$\star$})  & $10^{11}$ &  1  & $10^{11}$ &  $5.2\times 10^{-5}$ & $1.8\times 10^{-11}$  \\
\hline
\end{tabular}
\end{center}
\caption{\it \textbf{Majorana Leptogenesis:} BPs are corresponding to the star symbol points Black (BP 1), Red(BP 2), and Yellow(BP 3) represented in Fig.\ref{fig:maj_mn_vs_phi}. }
\label{table_Maj_BPs}
\end{table}

We observe that in Dirac case for the low scale of $M_{\zeta_1} (M_c)$ the BAU allowed parameter space is restricted to low scales of $\langle \varphi \rangle$ and the Dirac Leptogenesis allowed parameter space may leave hints in detector sensitivities like LISA, $\mu$ARES and DECIGO only. Whereas for the high scale of $M_{\zeta_1} (M_c)$ the Dirac Leptogenesis allowed parameter space may leave hints at various GW detectors like DECIGO, $\mu$ARES, LISA, ET, CE, SKA and CMB-lensing-future. In Majorana case for the low scale of $M_{N_1}$ the baryon asymmetry is inefficient to get the correct abundance, based on the mass ratio of RHNs, $M_{N_2}/M_{N_1}$. Whereas for high scale $M_{N_1}$ the BAU may achievable while leaving hints at DECIGO, $\mu$ARES, LISA, ET, CE and SKA. For both Dirac and Majorana cases the BPs corresponding to $\langle \varphi \rangle=\{10^{13}, 10^{12}, 10^{11}\}$ GeV show the high scale Leptogenesis may probe by the stochastic gravitational waves emanating from a network of cosmic strings.

\medskip

\section{Discussion and Conclusion}\label{sec:conclusion}

Thermal Dirac Leptogenesis through the Dirac seesaw mechanism gives an elegant and minimal explanation for two
outstanding puzzles in the standard model namely the generation of tiny neutrino masses and matter-antimatter asymmetry of the Universe involving no lepton number violation (LNV) on the whole \footnote{In B-L extended Dirac leptogenesis one may have LNV.}. Unfortunately, the scales of new physics, namely the seesaw scale and the Leptogenesis scale are naturally well beyond what we can directly test in terrestrial laboratory experiments. Given the fundamental nature of these puzzles, indirect cosmological tests of Dirac Leptogenesis are of great value and we propose cosmic strings as a powerful witness of the paradigm. We find that, if present, gravitational wave radiation from cosmic strings will carry the imprints of thermal Dirac leptogenesis. If there is U(1)$_{\rm B-L}$ survives below from GUT symmetry breaking scales to these scales to protect the right handed neutrino mass, then we show that cosmic strings appear through the breaking of this B-L symmetry predicting a spectrum of stochastic gravitational waves and, with future detectors being able to probe the entire mass range relevant to the paradigm of Dirac Leptogenesis. If cosmic strings GW spectrum is discovered and falls into the energy scales relevant for the Dirac seesaw mechanism and Leptogenesis, it would provide exciting hints of dynamics in the lepton sector at high energy scales. We showed a comparative analysis between such probes of GW in the context of Majorana and Dirac Leptogenesis within the B-L extension paradigm, and based on the energy scales of seesaw and Leptogenesis. We have shown, for instance, LISA detector will be able to probe the scale of seesaw  $\langle\varphi \rangle= 10^{11}$ GeV or higher while ET will be able to probe $\langle\varphi \rangle= 10^{12}$ GeV  or higher. Along with these, Figs.\,\ref{fig:lambda_vs_phi} and \ref{fig:maj_lambda_vs_phi} respectively also tell us about the scales of Dirac and Majorana seesaws, for which correct BAU can be satisfied by varying the Yukawas ($\text{Tr}[\lambda^{\dagger} \lambda]$) which control the CP-violation. It can be inferred that for the Majorana case, $\langle\varphi \rangle \geq 10^{12}$ GeV agree with observed BAU while below $\langle\varphi \rangle \leq 10^{12}$ GeV, the $Y_B \leq 8.66 \times 10^{-11}$ given $y_m = 1, \frac{M_{\rm N_2}}{M_{\rm N_1}} = 10^2$. This situation is however relaxed for the Dirac seesaw case, and the observed BAU can be satisfied for a wide range of values of $\langle\varphi \rangle$. Given the correlations between the scale of seesaw and scale of Leptogenesis, which are different for Dirac Leptogenesis and Majorana Leptogenesis, ET detector will be able to probe the scale of Dirac Leptogenesis $M_{\rm \zeta_1} \sim 10^{9}$ GeV or higher (see Fig.\,\ref{fig:mzeta1_vs_phi}\,), while the same for Majorana Leptogenesis $M_{N_{1}} \geq 10^{12}$ GeV within perturbative coupling limits (see Fig.\,\ref{fig:maj_mn_vs_phi}\,). LISA on the other hand will be able to probe even upto lower scales of Dirac Leptogenesis, $M_{\rm \zeta_1} \sim 5 \times 10^{8}$ GeV, and that of Majorana Leptogenesis $M_{N_{1}} \gtrsim 8\times 10^{11}$ GeV.

Sources of primordial gravitational waves, like those from topological defects like cosmic strings, is a useful search for new physics especially high energy scales like that of Leptogenesis which is otherwise very challenging to test, if not impossible, in laboratory experiments. Gravitational wave astronomy aspires to achieve precision measurements with the current and planned network of GW detectors worldwide that are orders of magnitude better than the current detectors. This new era will make the dream of testing theories related to explanations for neutrino mass generation and matter-antimatter asymmetry of the Universe a reality in near future.

\medskip

\acknowledgments
Authors thank Jan Heisig for several useful discussions on the topic and Angus Spalding for comments on the manuscript. This work was in part supported by the JSPS KAKENHI Grants Nos. JP23KF0289, and JP24K07027 (K.K.), and the MEXT KAKENHI Grants No. JP24H01825 (K.K.).


\appendix
\section{Appendix: Signal-to-noise ratio (SNR)}
\label{SNR}
 Gravitational Wave Interferometers actually measure displacements of detector arms in terms of a what is known in terms of dimensionless strain-noise $h_\text{GW}(f)$ which is related to the GW amplitude. This can be straight-forwardly converted into the corresponding  energy density \cite{Garcia-Bellido:2021zgu}
\begin{align}
    \Omega_\text{exp}(f) h^2 = \frac{2\pi^2 f^2}{3 H_0^2} h_\text{GW}(f)^2 h^2,
\end{align}
where $H_0 = h\times 100 \;\text{(km/s)}/\text{Mpc}$ is the Hubble expansion rate today. 
The signal-to-noise ratio (SNR) for a projected experimental sensitivity $\Omega_\text{exp}(f)h^2$ is then estimated in order to assess the detection probability of the primordial GW background originating from the global cosmic string background following the prescription~\cite{Thrane:2013oya,Caprini:2015zlo}
\begin{align}
     \text{SNR}\equiv \sqrt{\tau \int_{f_\text{min}}^{f_\text{max}} \text{d}f \left(\frac{ \Omega_\text{GW}(f) h^2}{\Omega_\text{exp}(f) h^2}\right)^2 } \label{eq:SNR},
\end{align}
where $h=0.7$ and  $\tau = 4\; \text{years}$ is the observation time. Usually this is chosen to be $\text{SNR}\geq 10$ as the detection threshold for each individual GW detector.

\vspace{10em}
\newpage
\newpage

\bibliography{ref}
\bibliographystyle{JHEP}

\end{document}